# Laser-Machined Ultra-High-Q Microrod Resonators for Nonlinear Optics

Pascal Del'Haye[1*], Scott A. Diddams[1], Scott B. Papp[1†]

[1]National Institute of Standards and Technology (NIST), Boulder, CO 80305, USA

**Optical whispering-gallery microresonators are useful tools in microphotonics, and non-linear optics at very low threshold powers. Here, we present details about the fabrication of ultra-high-Q whispering-gallery-mode resonators made by $CO_2$-laser lathe machining of fused-quartz rods. The resonators can be fabricated in less than one minute and the obtained optical quality factors exceed $Q = 10^9$. Demonstrated resonator diameters are in the range between 170 μm and 8 mm (free spectral ranges between 390 GHz and 8 GHz). Using these microresonators, a variety of optical nonlinearities are observed, including Raman scattering, Brillouin scattering and four-wave mixing.**

Since their inception nearly 25 years ago [1], optical whispering-gallery mode resonators [2-10] have had a major impact on many fields related to linear and non-linear optics. Optical microresonators are used in photonics technologies for laser stabilization [11-15], optical filtering and wavelength division multiplexing [16, 17], as well as in nonlinear optics for Raman lasers [18], frequency comb generators [19-28], and Brillouin lasers [29]. In addition, they have become useful tools in cavity quantum electrodynamics [30], and cavity optomechanics, where they couple mechanical motion with optical fields [31]. Several new types of microresonators for different applications have been developed during the last years. Most of them require complex processing and clean-room procedures [2, 3] or manual polishing steps for their production [5, 9]. Here we present fabrication details for ultra-high-Q whispering gallery mode resonators that are machined and shaped with a $CO_2$ laser by lathe turning of cylindrical glass rods. These resonators were recently introduced in the context of optical frequency comb generation and mechanical control of their mode spacing [32]. The method of resonator fabrication is extremely fast and versatile and reproducibly generates optical quality factors exceeding $5\times10^8$, and up to $1\times10^9$, which is comparable to the highest observed optical quality factors in other fused-silica resonators [33-35, 26]. Moreover, our technique allows control of both the fundamental diameter as well as the shape of the resonator sidewall by controlled ablation of material from the glass preform. In this way, we are able to fabricate resonators with major diameters ranging from 170 µm up to 8 mm and with sidewall curvature radii between 15 µm and 125 µm. With their ultra-high quality factors and fabrication times below 1 minute, these resonators are excellent tools for nonlinear optics experiments including low-threshold Raman scattering, Brillouin scattering and optical frequency comb generation via cascaded four-wave mixing.

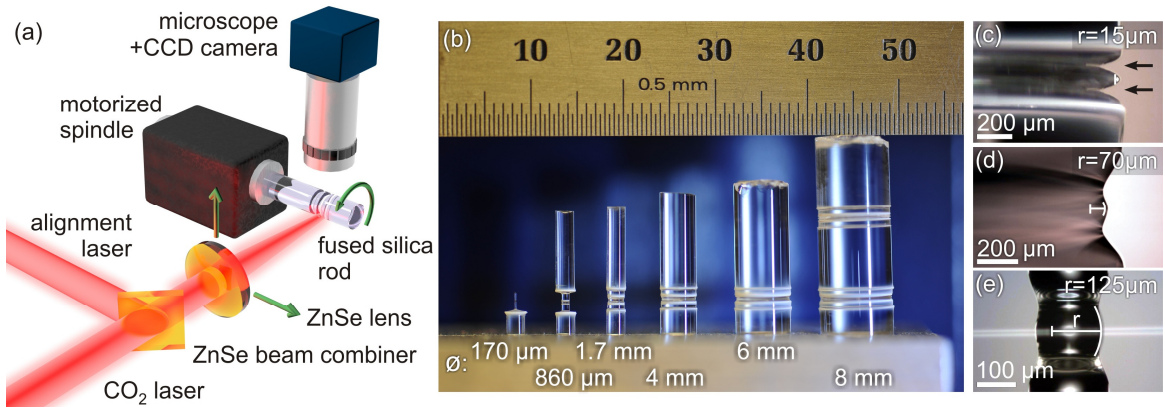

FIG. 1. (a) Setup for fabrication of ultra-high-Q microrod resonators. A fused-quartz rod is mounted on a motorized spindle and a focused $CO_2$ laser beam is used to shape a whispering-gallery mode resonator. The laser focus position can be controlled by a zinc selenide lens mounted on a motorized and computer-controlled translation stage. An additional collinear alignment laser at 633 nm is used for coarse positioning of the laser spot. (b) $CO_2$-laser-machined microrod resonators with diameters ranging from 170 µm up to 8 mm. (c)-(e) Different curvatures of the whispering-gallery side walls between 15 µm and 125 µm. The corresponding major diameters of the resonators are 2 mm in panel (c), 4 mm in (d) and 170 µm in (e).

Figure 1 shows the experimental setup for fabrication of whispering-gallery mode resonators from glass rod preforms. In our experiments, we used low-OH content (< 5 ppm in weight) fused-quartz rods with diameter of 2 mm, 4 mm, 6 mm and 8 mm for resonator fabrication. The preforms are mounted on a motor-driven ball-bearing spindle, as shown in Fig 1(a). Coarse alignment of the $CO_2$ laser beam is supported by a visible collinear alignment-laser that is superimposed with a zinc-selenide beam combiner. The laser beams are focused onto the fused-quartz rod with a zinc-selenide lens that is mounted on a computer-controlled translation stage with sub-micrometer resolution. In order to monitor the fabrication process, we use a microscope with CCD camera that is mounted on top of the glass rod. In addition to shaping the resonator, the $CO_2$ laser can also be used to change the diameter of the glass rod. This is done by turning the rod-preform at ~800 RPM and slowly approaching the laser focus from below while sweeping it back and forth along the rod axis (sweep speed 0.3 mm/s; $CO_2$-laser power ~30 W). All of the resonators in Fig 1(b) have been treated with this method in order to evaporate the surface layer and create a symmetric resonator with respect to the rotation axis of the motorized spindle. The first two resonators in Fig 1(b) have been shrunk more significantly, starting from a 2-mm-diameter rod down to 170 µm and 860 µm, respectively. The final diameter of the rod can be controlled to a level of ~10 µm, corresponding to a ~160 MHz control of the resonator's free spectral range in a 2-mm-diameter device. In a subsequent fabrication step, the actual whispering-gallery mode resonator is shaped by cutting two rings into the glass rod (arrows in Fig 1(c)). The remaining glass between the two rings confines the optical whispering-gallery mode. Moreover, the curvature of the whispering-gallery side walls can be controlled by adjusting the distance between the two rings. Fig 1(c)-(e) show different curvature radii between 15 µm and 125 µm radius, which affect the cross-section of the optical modes. Control of this curvature is important to increase the mode confinement as well as to change the dispersion of the resonator, e.g., for broadband optical frequency comb generation via four-wave mixing [36-38]. Note that the resonator in Fig 1(c) is shaped by directly focusing the $CO_2$-laser beam onto the glass rod, while the resonators in Fig 1(d),(e) are fabricated by placing the center of the focus slightly below the rod. In the case of the directly applied laser beam, the cutting process self-terminates when the heat transport through the surrounding glass inhibits deeper cutting. Focusing below the glass rods also leads to a self-terminating cutting after a sufficient amount of material is evaporated. Both processes create a smooth surface at the resonator sidewall, induced by the surface tension of the melted glass. Best results were obtained by applying the $CO_2$ laser for 3 seconds at each axial cutting position for a total of 20 iterations while the resonator is turning at ~200 RPM. The laser power in the fabrication process is adjusted to accommodate for the size of the preform and the distance between the cutting rings.

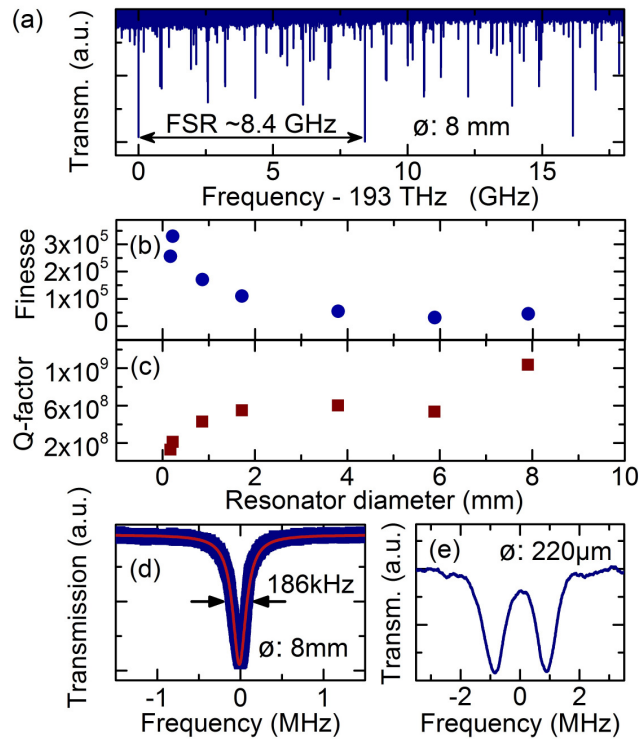

FIG. 2. (a) Mode spectrum of the 8-mm-diameter microrod resonator. The spectrum shows a diode laser sweep across two free spectral ranges of the resonator with approximately 30 different mode families. (b), (c) Finesse and optical quality factors of the resonators shown in Fig 1(b). The attained optical quality factors are between $2 \times 10^8$ and $1 \times 10^9$. (d) 186-kHz-wide mode ($Q \sim 1 \times 10^9$) in an 8-mm-diameter microrod resonator. (e) Mode splitting induced by coupling of clockwise and counter-clockwise modes in a smaller resonator (220 μm diameter).

The fabricated resonators are characterized for their optical quality and nonlinear properties with a coupling setup using a tapered optical fiber [39]. Figure 2(a) shows the mode spectrum of an 8-mm-diameter resonator obtained by sweeping an external cavity diode laser over ~2 free spectral ranges of the cavity. Around 30 different mode families are observed. The different mode families are polarization-dependent, and coupling to a certain mode family can be optimized by adjusting the polarization of the input light and the taper position. Figure 2(b) and 2(c) show the measured finesse and quality factor for resonators of sizes ranging from 170 µm diameter to 8 mm diameter. The obtained quality factors are above $Q=10^8$, with the highest quality factor exceeding $Q=10^9$ in the 8-mm-diameter device, which is most likely limited by OH-absorption [33, 40]. Based on the specified 5 ppm (weight) OH-content of the rod-preforms, we would expect material-loss-limited quality factors around $Q=2\times10^{10}$. The measured quality factors are expected to be slightly smaller as a result of an increased OH content during the fabrication process in air. In addition, the slightly reduced quality factors in smaller resonators are not yet limited by radiative whispering-gallery losses [1, 33] and could potentially be increased by further optimization of the fabrication parameters in order to reduce material and surface losses. Fig 2(d),(e) show measured mode profiles in an 8-mm-diameter and a 220-µm-diameter resonator. Mode splitting due to scattering induced coupling of clockwise and counter-clockwise modes [41] is only observed in small resonators with diameters of 220 µm and below. This suggests that the quality factors in larger devices are predominantly limited by light absorption in the material.

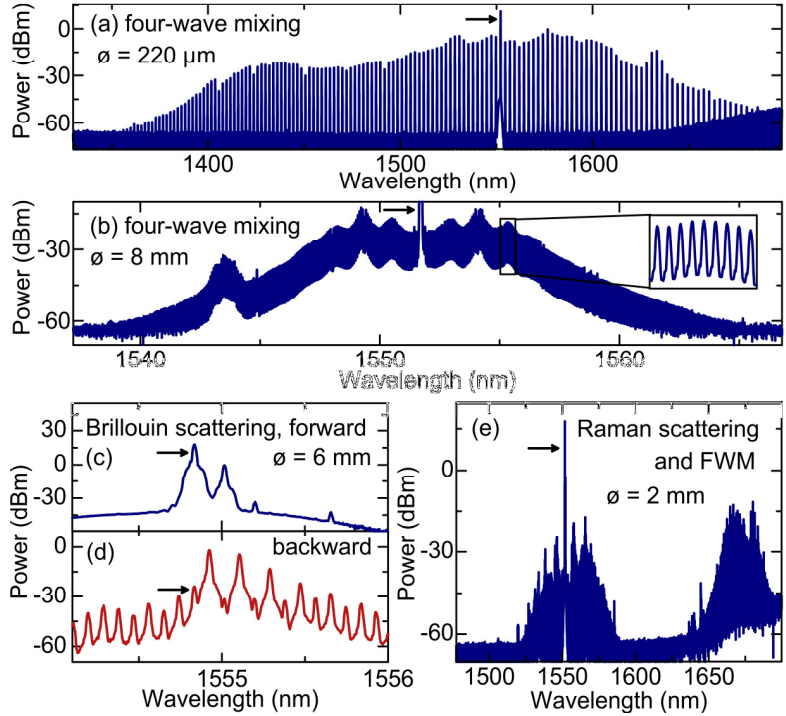

FIG. 3. Nonlinear optical effects in microrod resonators. Arrows in the figure show the position of the wavelength of the pump laser. (a) Cascaded four-wave mixing in a 220-µm-diameter resonator (mode spacing ~300 GHz). (b) Cascaded four-wave mixing in an 8-mm-diameter resonator (mode spacing ~8.4 GHz). (c),(d) Cascaded Brillouin scattering in a 6-mm-diameter resonator. The free spectral range of the resonator (~11 GHz) is close to the maximum of the Brillouin gain bandwidth. The forward direction of the tapered fiber output only shows even orders of Brillouin sidebands, while the backward direction is dominated by the odd orders of sidebands. (e) Raman scattering and four-wave mixing in a 2-mm-diameter resonator.

With optical quality factors exceeding $Q=10^8$, this type of microrod-resonator is an ideal candidate for nonlinear optics at low threshold powers. Figure 3 shows measurements of different nonlinear effects in resonators of different sizes at a launched pump power of ~100 mW. Four-wave mixing induced frequency comb generation can be observed with different mode spacings between 300 GHz (Fig. 3(a)) and 8.4 GHz (Fig. 3(b)) according to the resonator size. This four-wave mixing process is largely suppressed in the 6-mm-diameter resonator with a mode spacing of ~11 GHz, which is close to the maximum of the Brillouin shift in fused silica. Here, we observe several orders of Brillouin sidebands, as shown in Fig 3(c),(d). As previously reported in chip-based disk resonators [42], even order sidebands are observed in the forward direction and odd orders of sidebands are scattered in the backward direction with respect to the pump light. The backward-directed spectrum in Fig 3(d) has been obtained through a fiber-optic circulator at the input of the tapered optical fiber that was used for coupling. Fig 3(e) shows an example of a mixture of Raman scattering and four-wave mixing [43, 18] in a 2-mm-diameter microresonator. The threshold for Raman scattering and four-wave mixing in this device is around 1 mW and depends on the coupling conditions as well as the resonator geometry. Adjusting these parameters allows one to change the relative nonlinear threshold of these two effects.

In summary, we introduced ultra-high-Q whispering-gallery microresonators that are fabricated by $CO_2$-laser machining of cylindrical glass rods and characterized their optical properties for linear and nonlinear optical applications. The fabrication procedure of these resonators is exceedingly fast (~1 minute) and does not require expensive cleanroom facilities or time-consuming processing steps. We observed optical quality factors up to $Q=10^9$ and demonstrated resonator sizes between 200 µm and 8 mm diameter. In addition we have shown control of the curvature of the whispering gallery resonator sidewalls, which allows for control of modal confinement and resonator dispersion. The high quality factors enabled the observation of different nonlinear optical effects in these resonators. This fabrication method might be applied to a variety of glassy materials (in particular, glasses with high nonlinear coefficients) that have not previously been amenable to high-Q microresonator research.

**Acknowledgements:** This work is supported by NIST, the DARPA QuASAR program and NASA. PD thanks the Humboldt Foundation for support. This paper is a contribution of NIST and is not subject to copyright in the United States.

[*]pascal.delhaye@gmx.de

[†]scott.papp@nist.gov